\documentclass[aps,prl,preprint,superscriptaddress]{revtex4-2}

\usepackage{graphicx}
\usepackage{amsmath}
\usepackage{xcolor}
\usepackage{physics}
\usepackage[colorlinks=true,linkcolor=blue,urlcolor=blue,citecolor=blue]{hyperref}

\usepackage{bm}

\newcommand{\br}{\bm{r}}

\newcommand{\s}{_\mathrm{{\scriptscriptstyle S}}}
\newcommand{\h}{_\mathrm{{\scriptscriptstyle H}}}
\newcommand{\xc}{_\mathrm{{\scriptscriptstyle XC}}}

\newcommand{\ion}{_\mathrm{{\scriptscriptstyle ion}}}

\begin{document}


\title{Ab initio insights on the ultrafast strong-field dynamics of anatase TiO$_2$}


\author{Sruthil Lal S.B}
\affiliation{Department of Physics, Pondicherry University, R. V. Nagar, Kalapet, Puducherry, India}

\author{Lokamani}
\affiliation{Helmholtz-Zentrum Dresden-Rossendorf, 01328 Dresden, Germany}

\author{Kushal Ramakrishna}
\affiliation{Center for Advanced Systems Understanding, 02826 Görlitz, Germany}
\affiliation{Helmholtz-Zentrum Dresden-Rossendorf, 01328 Dresden, Germany}

\author{Attila Cangi}
\affiliation{Center for Advanced Systems Understanding, 02826 Görlitz, Germany}
\affiliation{Helmholtz-Zentrum Dresden-Rossendorf, 01328 Dresden, Germany}

\author{D Murali}
\affiliation{Indian Institute of Information Technology Design and Manufacturing (IIITDM), Kurnool, Andhra Pradesh, India}

\author{Matthias Posselt}
\affiliation{Helmholtz-Zentrum Dresden-Rossendorf, 01328 Dresden, Germany}

\author{Assa Aravindh Sasikala Devi}
\affiliation{ Nano and molecular systems research unit, P.O.Box 8000, FI-90014, University of Oulu, Oulu, Finland}

\author{Alok Sharan}
\email[]{aloksharan@gmail.com}
\affiliation{Department of Physics, Pondicherry University, R. V. Nagar, Kalapet, Puducherry, India}


\date{\today}

\begin{abstract}
    Electron dynamics of anatase TiO$_2$ under the influence of ultrashort and intense laser field is studied using the real-time time-dependent density functional theory (TDDFT). Our findings demonstrate the effectiveness of TDDFT calculations in modeling the electron dynamics of solids during ultrashort laser excitation, providing valuable insights for designing and optimizing nonlinear photonic devices. We analyze the perturbative and non-perturbative responses of TiO$_2$ to 30 fs laser pulses at 400 and 800 nm wavelengths, elucidating the underlying mechanisms. At 400 nm, ionization via single photon absorption dominates, even at very low intensities. At 800 nm, we observe ionization through two-photon absorption within the intensity range of $1\times10^{10}$ to $9\times10^{12}$ W/cm$^2$, with a transition from multiphoton to tunneling ionization occurring at $9\times10^{12}$ W/cm$^2$. We observe a sudden increase in energy and the number of excited electrons beyond $1\times10^{13}$ W/cm$^2$, leading to their saturation and subsequent laser-induced damage. We estimate the damage threshold of TiO$_2$ for 800 nm to be 0.1 J/cm$^2$. In the perturbative regime, induced currents exhibit a phase shift proportional to the peak intensity of the laser pulse. This phase shift is attributed to the intensity-dependent changes in the number of free carriers, indicative of the optical Kerr effect. Leveraging the linear dependence of phase shift on peak intensities, we estimate the nonlinear refractive index ($n_2$) of TiO$_2$ to be $3.54\times10^{-11}$ cm$^2$/W.
\end{abstract}


\maketitle

\section{Introduction}
Time-dependent density-functional theory (TDDFT)~\cite{runge1984density} describes the quantum dynamics of electrons under the influence of a time-dependent external potential~\cite{bertsch2000real,yabana2012timedependent,yamada2019energy,uemoto2019nonlinear,uemoto2021first,hashmi2022nonlinear}. TDDFT calculations are used to study ultrashort laser-matter interactions including high-harmonic generation (HHG)~\cite{tancogne2017impact,floss2018abinitio,otobe2012first}, nonlinear current injection~\cite{wachter2014ab,wachter2015controlling}, formation of Floquet-Bloch states~\cite{de2016monitoring,hubener2017creating}, laser ablation~\cite{sato2015timedependent,sato2016firstprinciples}. TDDFT computations have also been utilized to distinguish between the purely electronic and phononic contribution to non-equilibrium dynamics in metals caused by lasers~\cite{niedermayr2022few}. Furthermore, TDDFT has been applied to study the influence of the laser pulse widths in the formation of Nitrogen-Vacancy centers in the diamond lattice~\cite{shimotsuma2023formation}.

The strong-field response in solids has become an area of renewed interest due to recent experimental evidence that dielectrics can survive electric fields approaching their critical fields when exposed to laser pulses shorter than the electronic relaxation time scales~\cite{ghimire2011observation, schultze2013controlling, schubert2014sub}.  Initiating, driving, and probing the nonlinear electron dynamics in crystalline materials is now possible with optical sub-cycle resolutions, opening the door for optical field-effect devices operating within single optical cycles and petahertz signal processing~\cite{durach2011predicted, schiffrin2013optical, lucchini2016attosecond, schultze2013controlling, apalkov2013metal, schultze2014attosecond, schubert2014sub, sommer2016attosecond, sederberg2020attosecond}. For instance, a reversible energy exchange at sub-30-attosecond timescales was observed in fused silica by Sommer et al.~\cite{sommer2016attosecond}. Under the influence of strong electric fields, it has been shown that the AC conductivity of fused silica increases by 18 orders of magnitude within one femtosecond~\cite{schiffrin2013optical} and is completely reversible. TDDFT calculations have shown that electron tunneling is the fundamental mechanism of carrier injection in silica under few-cycle extreme ultraviolet (XUV) illumination~\cite{schultze2014attosecond}. Materials undergo dynamic metallization~\cite{durach2010metallization,durach2011predicted,apalkov2012theory,kwon2016semimetallization} when irradiated with optical pulses of amplitude as large as 1 V/{\AA}. This observation was also supported by TDDFT calculations~\cite{wachter2014ab}.

TDDFT calculations of ultrashort laser-induced electron dynamics for nonlinear photonic applications have so far been focused on Si~\cite{schultze2014attosecond,sato2015timedependent}, SiO$_2$~\cite{kwon2016semimetallization,yamada2019energy,sato2015timedependent}, linear carbon chain~\cite{su2016first}, diamond~\cite{zhang2017manipulation,bertsch2000real,kazempour2021transient,wang2013nonlinear}, phospherene~\cite{shinde2018nonlinear}, MoS$_2$~\cite{su2017electron}. Titanium dioxide (TiO$_2$), commonly used as a saturable absorber in passively Q-switched fiber lasers~\cite{rusdi2017titanium,ahmad2015c}, has great potential for enabling nonlinear photonics. The nonlinear optical response of anatase TiO$_2$ has a typical recovery period of approximately 1.5 ps~\cite{elim2003ultrafast}. The nonlinear index ($n_2$) of bulk and thin-film of TiO$_2$ ranges from 0.8--3$\times 10^{-14}$ cm$^2$/W~\cite{evans2012mixed,hashimoto1994sol,castillo2010ultrafast}, which is greater than the nonlinear index of silica fiber (2.48$\times$ 10$^{-16}$ cm$^2$/W~\cite{milam1998review}). Moreover, the two-photon absorption of TiO$_2$ at 800 nm is minimal, making it ideal for waveguides operating near 800 nm~\cite{reshef2015polycrystalline}. TiO$_2$ can be formed at low temperatures ($<$400~$^{\circ}$C) and offers advantages over silicon nitride with its higher refractive index (2.4 vs. 2.0) and more than three times stronger Kerr non-linearity~\cite{evans2012mixed,bradley2012submicrometer,choy2012integrated}. These properties enable back-end integration with silicon micro photonic devices. Existing estimates of $n_2$ of TiO$_2$ are either from femtosecond z-scan measurements or by fitting the nonlinear pulse propagation simulations (based on the nonlinear Schrodinger equation) to the experimental data~\cite{evans2013spectral}. A systematic analysis of ultrafast nonlinear optical interactions in TiO$_2$ from a microscopic perspective has yet to be explored. 

This study uses first-principle simulations to examine the microscopic electron dynamics of crystalline anatase TiO$_2$ modulated by an ultrashort and intense laser fields. We employ TDDFT calculations as implemented in the software package OCTOPUS~\cite{tancogne2020octopus}. We explore the response of anatase TiO$_2$ to 800 nm and 400 nm laser pulses with intensities spanning from the perturbative to strong-field regimes (non-perturbative). Different regimes of nonlinear interactions with the external electric field are characterized, and various underlying mechanisms are analyzed. The evolution of photoinduced current and energy transfer during the interaction is studied. We determine the nonlinear refractive index and optical damage threshold of anatase TiO$_2$, and our results are in excellent agreement with previously reported experimental data.

The paper is organized as follows. Section~\ref{sec:comp_methods} describes the computational methods employed for determining the photoinduced current and the energy dynamics of TiO$_2$. The results and analysis of our study are discussed in Section~\ref{sec:results}, where we also compare them with the existing experimental data. We conclude the paper with a summary in Sec.~\ref{sec:conclusion}.

\section{Computational Methods}
\label{sec:comp_methods}
\subsection{Time-dependent Density Functional Theory}
The electron dynamics in a unit cell of a periodic crystal driven by a time-dependent electric field $\boldsymbol{E}(\br,t)$ is described in terms of the time-dependent Kohn-Sham (KS) equations
\begin{equation}
    \label{eq:tdks}
	i\frac{\partial}{\partial t}u_{n,k}(\textbf{r}, t) = \left\{\frac{1}{2}\left[\textbf{p} + \textbf{A}\s(\br, t) \right]^2 + v\s(\br, t) \right\} u_{n,k}(\br, t),
\end{equation}
where $u_{n,k}(\br, t)$ denotes KS orbitals with the band index $n$, the electron wave vector $k$, and $v\s[n](\br, t) = v\ion(\br, t) + v\h[n](\br, t) + v\xc[n](\br, t)$ is the KS potential with $v\ion$ denoting the external ionic potential, $v\h$ the Hartree potential, and $v\xc$ the exchange-correlation (XC) potential. Furthermore, $\textbf{p}$ is the momentum operator and $\textbf{A}\s(\br, t)=\textbf{A}(\br, t) + \textbf{A}\xc(\br, t)$ is the vector potential composed of the applied vector potential $\textbf{A}(\br, t)$ and an XC contribution $\textbf{A}\xc(\br, t)$~\cite{vignale1996current}.  The applied vector potential represents an applied electromagnetic field, such as a laser pulse, and is related to the applied electric field by $\boldsymbol{E}(\br, t) = -(1/c)[\partial \textbf{A}(\br, t)/\partial t]$. Note that the laser pulse can be treated as spatially uniform $\boldsymbol{E}(t)$ under the dipole approximation. Solving the time-dependent KS equations with the exact XC potential and XC vector potential yields the exact time-dependent electron density $n(\br, t) = \sum_{n,\textbf{k}}^{occ} u^*_{n,k}(\br, t)\, u_{n,k}(\br, t)$.
However, in practice approximations are used, e.g., a particular approximation is used to express the XC potential~\cite{rappoport2005approximate}, the adiabatic approximation is applied, and the $A\xc$ is often neglected.
We follow the general practice by applying these approximations as detailed below. Note that we adopt Hartree atomic units, i.e., $\hbar = e = m= 1$. 

Another useful quantity is the microscopic current density which is determined in the outlined framework as
\begin{equation}  
        \label{eq:current_density}
	\boldsymbol{j}(\br, t) = \sum_{n\textbf{k}}^{occ} \frac{1}{2}\left[u_{n,k}^{*}(\br, t) \left(\textbf{p} +\boldsymbol{A}(t) \right) u_{n,k}(\br, t)\right] ,
\end{equation}
where the summation runs over the occupied bands. The macroscopic current density $J(t)$ along the laser polarization direction $\boldsymbol{E_0}$ is obtained by averaging $\boldsymbol{j}(\br, t)$ over the unit cell with volume $\Omega$, 
\begin{equation}
    J(t) = \frac{1}{\Omega}\int_{\Omega}d^3\boldsymbol{r}\boldsymbol{j}(\br, t)\cdot  \boldsymbol{E_0}/\lvert\boldsymbol{E_0}\rvert. 
\end{equation}

The polarization density corresponding to $J(t)$ is $P(t) = \int_{0}^{t} J(t')dt'$. The time-resolved energy density $ W(t)$ transferred between the field and the material is evaluated by 
\begin{equation}
	\label{eq:energy_transfer}
	 W(t) = \int_{-\infty}^t  dt'\  \boldsymbol{E}(t) \cdot \boldsymbol{J}(t).
\end{equation}
Its resultant value at the end of the laser pulse $ W(t\to\infty)$ determines the total amount of energy dissipated during the light-matter interaction. The number of electrons excited from the valence band to the conduction band $N_{exc}$ per unit cell is calculated using~\cite{otobe2008first}
\begin{equation}
	\label{eq:nex}
	N_{exc}(t) = \sum_{n,n',\boldsymbol{k}} \left(\delta_{nn'\textbf{k}} - \mid \bra{u_{n,k}(0)} \ket{u_{n',k}(t)} \mid^2 \right)\ .
\end{equation}
Here $u_{n,k}(0)$ is the KS orbital of the initial state, $u_{n'k}(t)$ is the time-dependent KS orbital, and $\delta$ is the Kronecker delta function. 

We use the real-space, real-time code Octopus~\cite{tancogne2020octopus} to carry out the TDDFT calculations. The laser-induced dynamics of valence electrons are calculated in a unit cell of anatase TiO$_2$. Anatase TiO$_2$ crystallizes with a tetragonal unit cell having a lattice spacing $a = 3.97 $ {\AA} and $c/a = 2.52 $. We treat the interaction $v\ion(\br, t)$ between valence electrons and the ionic core by the Kleinman-Bylander pseudopotential~\cite{kleinman1982efficacious}. The generalized gradient approximation (GGA) based on the Perdew-Burke-Ernzerhof functional (PBE)~\cite{perdew1996generalized} is employed for the XC potential. KS orbitals are represented on the discretized real-space grid with $\Delta x = \Delta y = 0.12 $ {\AA} and $\Delta z = 0.20 $ {\AA}. It is equivalent to a plane-wave cut-off at 900 eV. The time-dependent KS equations are solved on a uniform grid with $\approx29000$ grid points. The Brillouin zone is uniformly sampled by $12\times 12 \times4$ Monkhorst-Pack grids~\cite{monkhorst1976special}. The discretization consists of 363 symmetry-reduced k-points for $x$ polarized light. With this setup, the system's total energy converges to within 1~meV. 

First, the ground state of TiO$_2$ is calculated, which will be used as the initial state for the time-dependent calculations. We then time-propagate the KS orbital by solving Eq.~(\ref{eq:tdks}) in the time domain. The time evolution is calculated with the approximated enforced time-reversal symmetry (AETRS)~\cite{castro2004propagators} as the time-evolution propagator with a time step $\Delta  t= 0.02\ a.u.$. The total simulation duration is 30 fs (1240 atomic units with a step size of 0.02 a.u., i.e., $\approx$64000 time steps). Note that, during the time evolution, ions are at their equilibrium positions in the ground state. Furthermore, the adiabatic approximation~\cite{thiele2008adiabatic} is used which means that the time dependence of the XC potential is approximated by evaluating a ground-state XC functional at the time-dependent density.

We calculate the response of TiO$_2$ to a linearly polarized laser pulse, which is represented by the spatially-uniform electric field through the corresponding vector potential 
\begin{equation}
    \label{eq:laser_profile}
    \textbf{A}(t) = \frac{\boldsymbol{E_0}}{\omega}\exp\left[\frac{-(t-t_0)^2}{2\tau_0^2}\right]\cos(\omega t)\,,
\end{equation}
where $\omega$ is the central frequency of the laser pulse and $\boldsymbol{E_0}$ is the amplitude of the time-dependent electric field $\boldsymbol{E}(t)$, which is related to the laser peak intensity $I_0 = c\lvert \boldsymbol{E_0}\rvert^2/8\pi$.

\section{Results}
\label{sec:results} 
The following section presents the electron dynamics of crystalline anatase TiO$_2$ excited by 800 nm and 400 nm laser pulses represented by Eq.~(\ref{eq:laser_profile}). The duration of the pulse is set to T =30 fs ($\approx $12 fs at the FWHM of the envelope.), while the amplitude of the pulse is varied from $10^{7}$ to $10^{16}$ W/cm$^2$. The laser field is polarized along the $x$-axis. 
\subsection{Energy Transfer Dynamics}
\label{sec:energy_transfer}
\begin{figure*}
	\centering{}
	\includegraphics[width=0.95\textwidth, keepaspectratio]{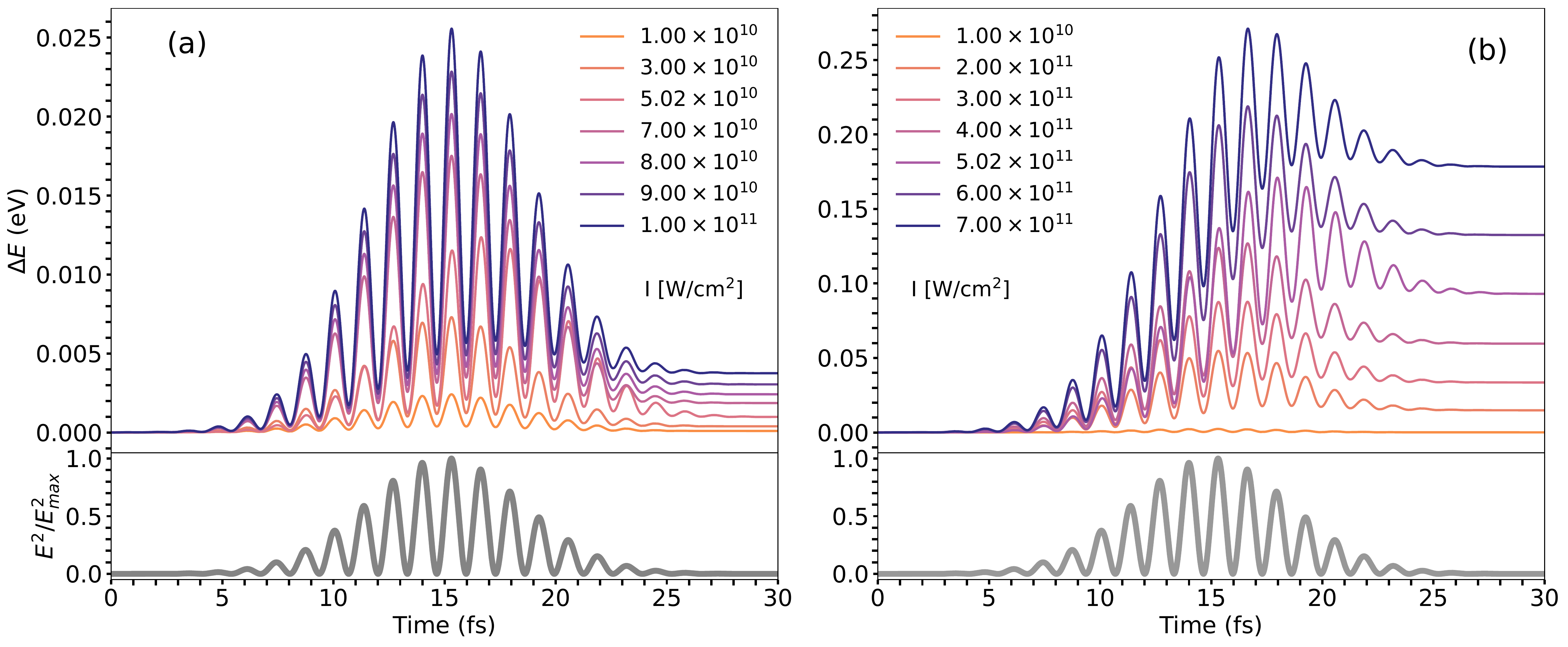}
    \caption{\label{fig:energy_transfer}Time-dependent energy exchanged between anatase TiO$_2$ and 30 fs pulses at 800 nm is given for different peak intensities: Panel (a) represents the non-resonant virtual energy transfer where the transferred energy oscillates synchronously $\boldsymbol{E}^2(t)$ (bottom panel). This occurs for intensities below $1\times 10^{11}$ W/cm$^2$. Panel (b) shows the energy exchange via resonant two-photon absorption for intensities ranging from $2 \times 10^{11}$ to $7\times 10^{11}$ W/cm$^2$. For panels (a) and (b) the dynamics at $1\times 10^{10}$ W/cm$^2$ is shown as the reference.}
\end{figure*}
The energy transferred from the applied electric field to anatase TiO$_2$ is evaluated by Eq.~(\ref{eq:energy_transfer}). Fig.~\ref{fig:energy_transfer} shows the resultant energy dynamics for incident laser pulses at 800 nm ($\hbar \omega = 1.55$ eV) with different peak intensities. The central frequency of the pulse corresponds to energy lower than the direct gap (2.25 eV)~\cite{S_fig1}, leading to two general types of temporal energy transfer profiles. The first type is non-resonant excitation. The transferred energy, in this case, oscillates synchronously with the $\boldsymbol{E}^2(t)$, and the system almost returns to the ground state at the end of the pulse. This represents a virtual energy transfer from the laser pulse to the electrons. Such dynamics is observed in Fig.~\ref{fig:energy_transfer}(a) for peak intensities from $1 \times 10^{10}$ to $1\times 10^{11}$ W/cm$^2$. This behavior is typical when the frequency is below the bandgap, and the intensity is very low. 

The second kind of response is resonant excitation, where, along with the virtual oscillations, the transferred energy gradually increases during the pulse and persists beyond the pulse width. Given the pulse energy is below the bandgap, this occurs when the field is strong enough to induce real excitation through multi-photon absorption. Fig.~\ref{fig:energy_transfer}(b) illustrates the energy transfer $W(t)$ for this scenario, observed for intensities ranging from $2 \times 10^{11}$ to $7\times 10^{11}$ W/cm$^2$. The permanent energy transfer is related to creating electron-hole pairs, corresponding to the population transfer from valence bands to conduction bands.

\begin{figure}
	\centering{}
	\includegraphics[width=0.8\textwidth, keepaspectratio]{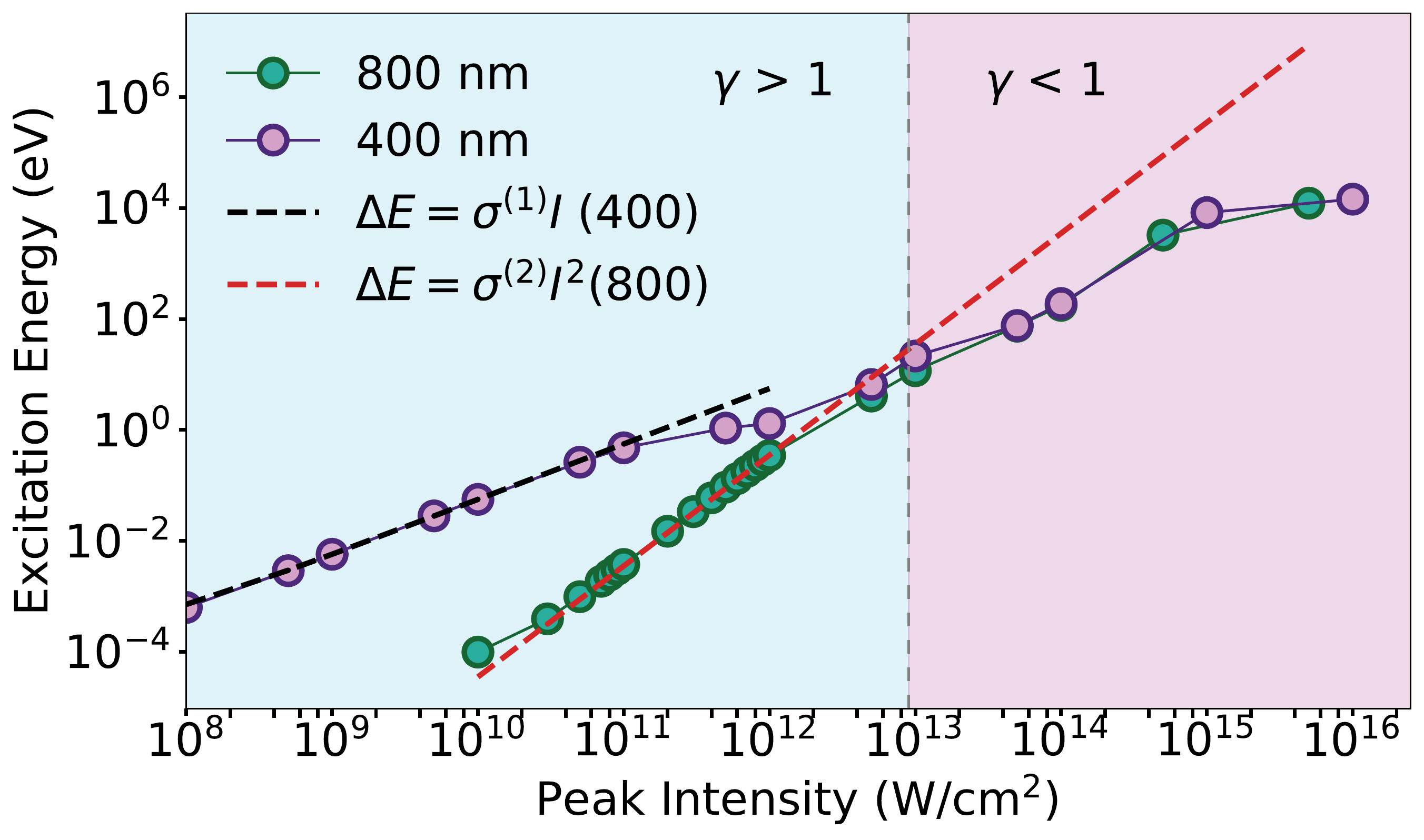}
    \caption{\label{fig:residual_energy}The peak laser intensity-dependence on energy absorbed in anatase TiO$_2$ crystal pumped by 800 nm and 400 nm laser pulses. Energy exchange at 400 nm is predominantly through single photon absorption for all intensities up to $\approx 5\times 10^{12}$ W/cm$^2$. For 800 nm pulse, no energy is exchanged until peak intensity reaches $\geq 1\times 10^{10}$, while two-photon absorption becomes dominant for intensities ranging from $ 1\times 10^{10}$ W/cm$^2$ upto $1\times 10^{12}$ W/cm$^2$. Typical intensity ranges of 800 nm pulses over which the multi-photon absorption ($\gamma >1$) or tunneling ionization ($\gamma <1$) becomes the dominant process are highlighted.}
\end{figure}

Figure~\ref{fig:residual_energy} illustrates the residual excitation energy in anatase TiO$_2$ after interacting with 800 and 400 nm laser pulses at different peak intensities. The energy absorbed at a wavelength of 400 nm is directly related to the intensity of the light and can be accurately described by the equation $\sigma^{(1)}I$ governing single-photon absorption, where $\sigma^{(1)}$ is a constant coefficient. This relationship holds true for light intensities lower than $1\times 10^{11}$ W/cm$^2$. This linear absorption behavior at 400 nm is expected since the single photon (3.10 eV) bridges the direct gap of anatase TiO$_2$. Conversely, single photon absorption below the direct bandgap is unlikely at 800 nm, and hence, no permanent energy transfer for intensities below $1\times 10^{10}$ W/cm$^2$. As the intensity increases from $ 1\times 10^{10}$ W/cm$^2$ upto $1\times 10^{12}$ W/cm$^2$, the deposited energy increases and closely follows a quadratic dependence $\sigma^{(2)}I^2$ on intensity (Fig.~\ref{fig:residual_energy}). 

At approximately $1\times 10^{13}$ W/cm$^2$ intensity, the excitation energies of 400 nm and 800 nm wavelengths combine to form a single curve. Below the intersection point, the excitation energy displays a perturbative behavior that can be effectively modeled by $I^N$, where $I$ represents the laser intensity, and $N$ corresponds to the number of photons required to exceed the bandgap energy. At intensities above the intersection region, the excitation energy is independent of laser frequency, and the curve's slope decreases compared to the region below the intersection~\cite{yamada2019energy}. It suggests that there is a saturation-like behavior occurring in the material's response. The similarity of the number density of excited electrons for both 800 nm and 400 nm beyond $\sim 10^{13}$ W/cm$^2$ also indicates the saturation effects~\cite{S_fig2}. For intensities higher than $1\times 10^{14}$ W/cm$^2$, the energy transfer exhibits an abrupt increase, indicating the onset of material laser-induced dielectric breakdown, as outlined in Sec.~\ref{sec:optical_damage}

\begin{figure}
	\centering{}
	\includegraphics[width=0.8\textwidth, keepaspectratio]{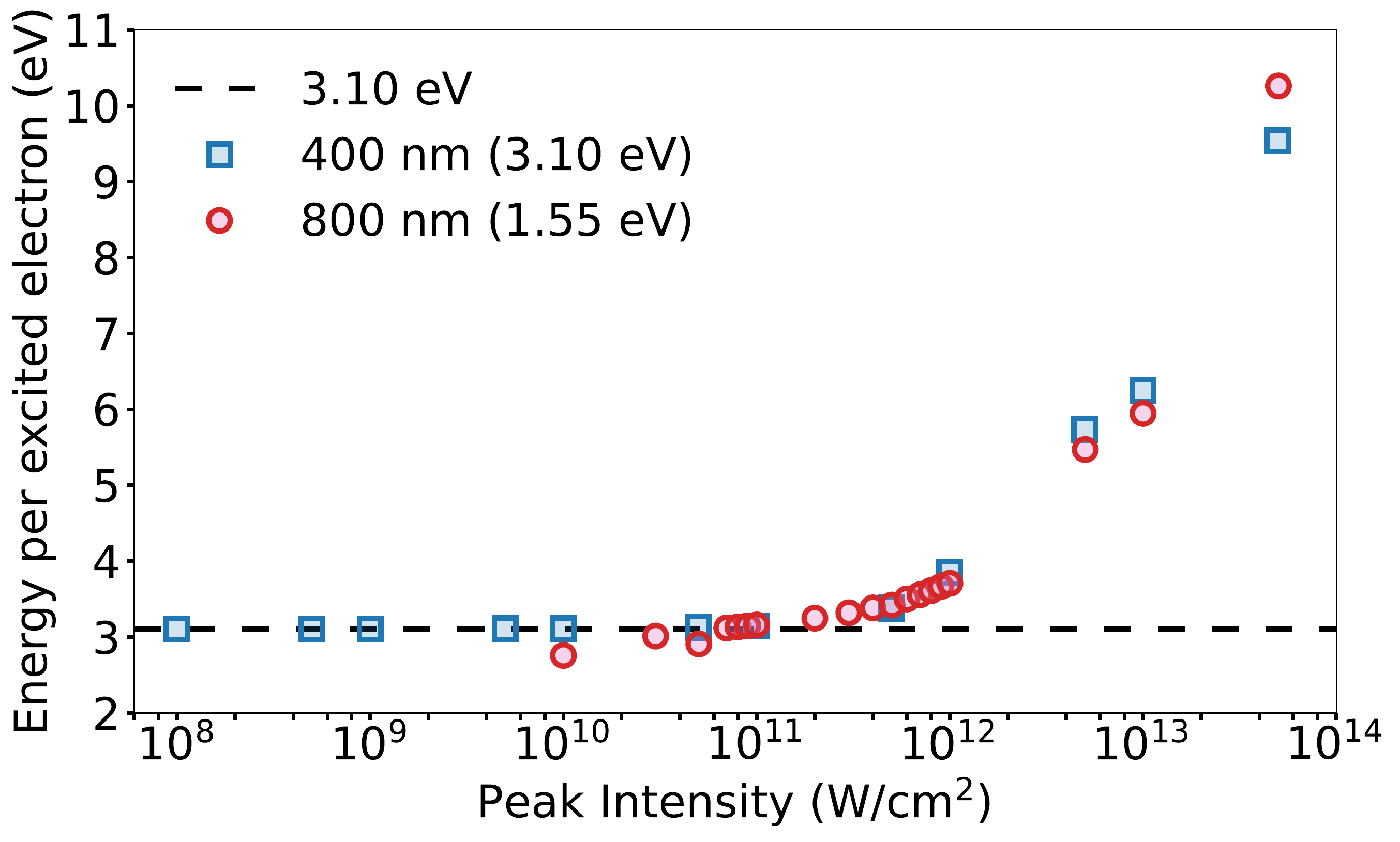}
    \caption{\label{fig:energy_per_electron}The energy of the excited electron ($E^{res}_{exc}$) at 800 nm and 400 nm beyond the pulse duration. At 400 nm, $E^{res}_{exc}$ is 3.10 eV for intensities up to $\approx 5\times 10^{12}$ W/cm$^2$, indicating single photon absorption. At 800 nm, absorption is unlikely for $I_0 \leq 1\times 10^{10}$ W/cm$^2$. For intensities ranging from $ 1\times 10^{10}$ W/cm$^2$ to $1\times 10^{12}$ W/cm$^2$ energy per electron lies on the two-photon absorption energy (3.10 eV). $E^{res}_{exc}$ gradually increases and, becomes frequency independent for intensities larger than $\approx 1\times 10^{12}$ W/cm$^2$, potentially due to higher-order multiphoton absorption and secondary excitation of excited electrons}
\end{figure}

Next, we analyze the energy of excited electrons at 800 nm and 400 nm beyond the pulse duration. The residual energy per excited electron ($E^{res}_{exc}$) is obtained by dividing the energy (Fig.~\ref{fig:energy_transfer}) by the number of excited electrons~\cite{S_fig2} at their saturation values. The results are shown in Fig.~\ref{fig:energy_per_electron}. At 400 nm, $E^{res}_{exc}$ is approximately 3.10 eV for intensities up to $\approx 1\times 10^{12}$ W/cm$^2$, indicating single photon absorption. At 800 nm, no excited electrons are observed until the intensity reaches $1\times 10^{10}$ W/cm$^2$. However, it approaches twice the photon energy (3.10 eV) for intensities ranging from $ 1\times 10^{10}$ W/cm$^2$ to $1\times 10^{12}$ W/cm$^2$, indicating ionization by two-photon absorption. $E^{res}_{exc}$ gradually increases above 3.10 eV reference line for intensities larger than $\approx 1\times 10^{12}$ W/cm$^2$  in Fig.~\ref{fig:energy_per_electron}, potentially due to higher-order multiphoton absorption and secondary excitation of excited electrons~\cite{keldysh1965ionization, reiss1980effect}.

The Keldysh parameter, denoted by $\gamma$, serves as an approximate measure to determine the type of strong field ionization~\cite{keldysh1965ionization}. The Keldysh parameter for the interaction of a laser pulse of frequency $\omega$ and field amplitude $E_0$ with a material of energy gap $\Delta$ is given by 
\begin{equation}
	\gamma = \frac{\omega\sqrt{m\Delta}}{eE_0} ,
\end{equation}
where $F[V/cm] = 27.44 \sqrt{I[W/cm^2]}$, $I$ is the peak intensity of the laser pulse and $e$ and $m$ are the charge and mass of electron, respectively. The condition $\gamma >1$ represents multi-photon ionization being the primary mechanism of ionization whereas $\gamma <1$ indicates that tunneling ionization dominates. As the intensity of the laser pulse increases, a transition from multi-photon absorption to tunneling ionization can be observed. The Keldysh parameter at 800 nm is calculated at different peak intensities. Based on the value of $\gamma$, the intensities over which multiphoton or tunneling ionization dominate are highlighted in Fig.~\ref{fig:residual_energy}. When $I \approx 9\times 10^{12}$ W/cm$^2$, the Keldysh parameter assumes a value of 1. It indicates that, for 800 nm, below an intensity of $9\times 10^{12}$ W/cm$^2$, the ionization is predominantly via multiphoton absorption and tunneling ionization above it.

\subsection{Saturation of photo-induced current at 400 nm}
\label{sec:current_saturation}
\begin{figure}
	\centering{}
	\includegraphics[width=0.7\textwidth, keepaspectratio]{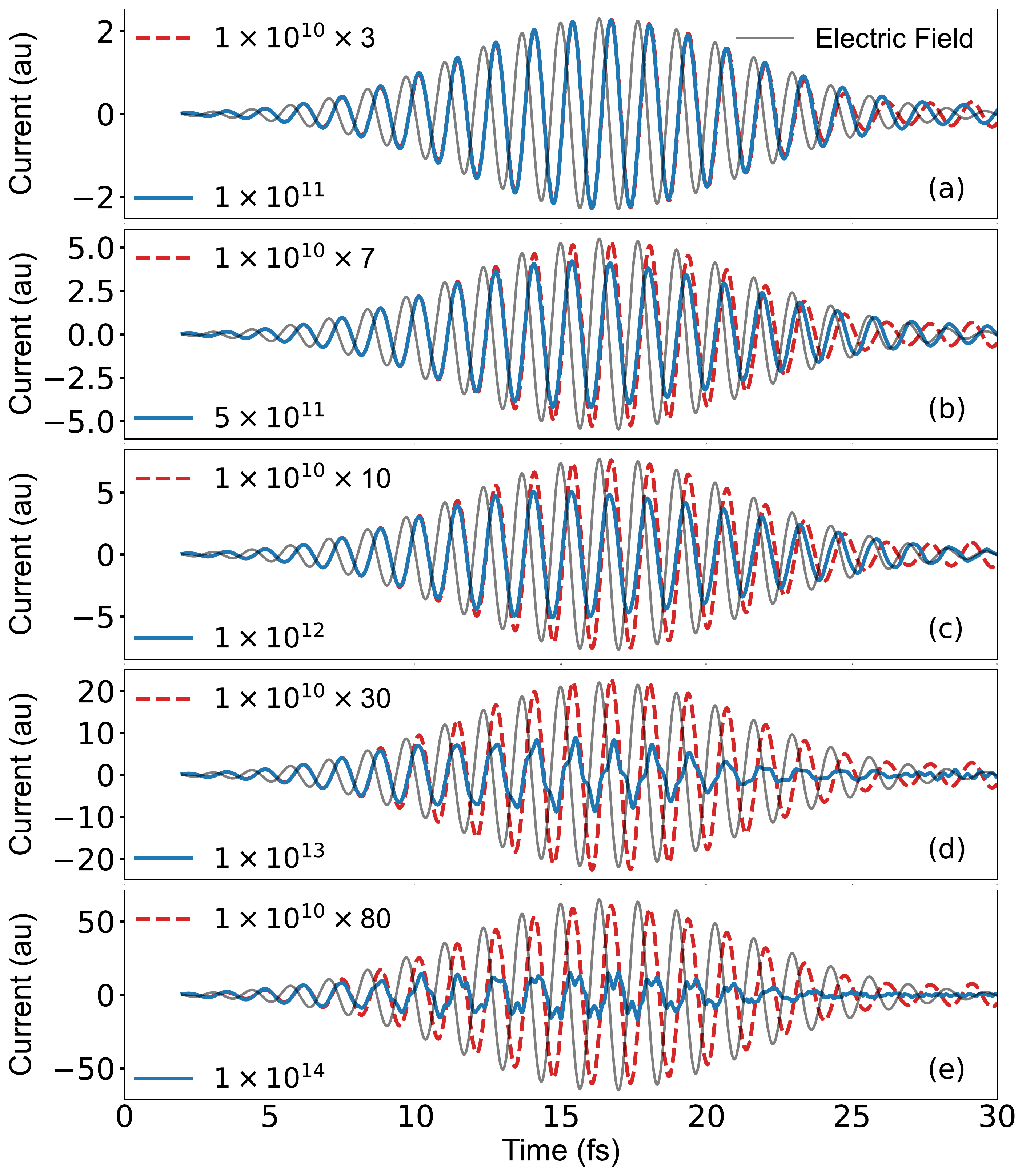}
    \caption{\label{fig:current_saturation}Current profiles for 400 nm laser pulses of total duration 30 fs showing the saturation of current as the peak intensity is increased from $I_0 = 1\times 10^{10}$ W/cm$^2$ to $I_0 = 1\times 10^{14}$ W/cm$^2$. This is a nonlinear optical effect occurring because of ground state bleaching due to linear absorption at 400 nm.}
\end{figure}

Figure~\ref{fig:current_saturation} shows the induced current for a laser pulse at 400 nm and a pulse duration of 30 fs with peak intensities ranging from $I_0 = 1\times 10^{10}$ W/cm$^2$ to $I_0 = 1\times 10^{14}$ W/cm$^2$. We take the current profile at $I_0 = 1\times 10^{10}$ W/cm$^2$ as the reference (weak) current to discuss the dynamics. In Figs.~\ref{fig:current_saturation} (a-e), the reference current is multiplied by a suitable factor so that the difference between currents at weak and strong field strengths indicates the nonlinear interaction. When the response is linear, the currents for weak and strong intensities will coincide and show similar profiles. 

In Fig.~\ref{fig:current_saturation}(a), the temporal evolution of the current at $I_0 = 1\times 10^{11}$ W/cm$^2$ follows the driving laser field, and it coincides with the reference current, indicating a linear response. The response is dielectric-like: the current is $\pi/2$ phase shifted with the electric field $\boldsymbol{E}(t)$. For $I_0 > 1\times 10^{11}$ W/cm$^2$ (Fig.~\ref{fig:current_saturation}(b-e)), the induced current is initially very close to that of the reference current. However, as the electric field of the pulse increases, the induced current gradually becomes weaker than expected from the linear response. This nonlinear effect of suppression of induced current occurs due to the bleaching of valence band electrons by absorption at 400 nm~\cite{hashmi2022nonlinear,uemoto2019nonlinear}. The majority of valence electrons are already excited, and the conduction bands are mostly filled, resulting in the suppression of further electron excitation. 
Additionally, because the frequency of the applied laser pulse is higher than the bandgap value, a significant current remains after the incident pulse has ended.

\subsection{The Nonlinear Refractive Index Change}
\label{sec:refractive_index}
\begin{figure}
	\centering{}
	\includegraphics[width=0.8\textwidth, keepaspectratio]{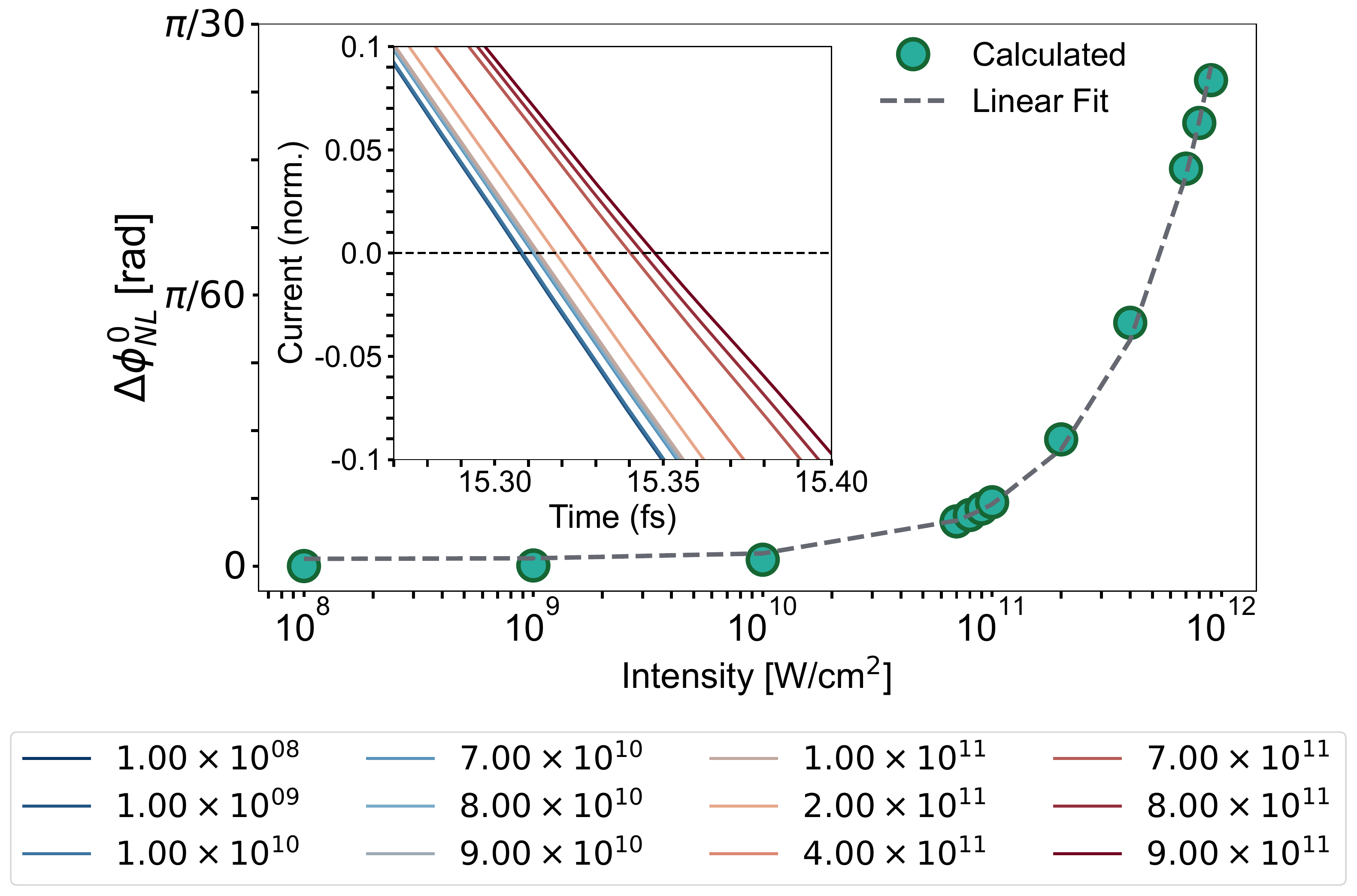}
    \caption{\label{fig:phase_shift_n2}The intensity scaling of phase shift of light-induced current at 800 nm is shown for different intensities. Phase shift is expressed by taking the current at $1\times 10^{8}$ W/cm$^2$ as the reference. The phase shift is determined from the temporal shift of the induced current calculated at the zero-crossing after the peak of the pulse ($\Delta \phi_{NL}^0$), as illustrated in the supplemental Fig.~\cite{S_fig3}. In the inset,  the induced current in the region close to the zero-crossing is zoomed in, highlighting the temporal shift. The phase shift can be related to the optical Kerr effect, according to which the increase in phase shift can be described as a linear rise with intensity $\Delta \phi_{NL}^0 = m\times I_0$. From the value of $m$ obtained from the figure, the nonlinear refractive index  $n_2 = 3.54 \times 10^{-11}$ cm$^2$/W is extracted. }
\end{figure}

The phase shift of light-induced current at 800 nm is depicted for various intensities in Fig.~\ref{fig:phase_shift_n2} with the current at $1\times 10^{8}$ W/cm$^2$ taken as the reference. For a pulse of given peak intensity, the induced current in the initial part of the pulse is in phase with the reference current. However, as the electric field of the pulse increases, the induced current starts accumulating a phase shift. The accumulated phase shift calculated from the temporal shift at the zero-crossing after the peak of the pulse ($\Delta \phi_{NL}^0$)~\cite{S_fig3} in Fig.~\ref{fig:phase_shift_n2} increases as the peak intensity is increased. The phase shift can be related to the optical Kerr effect where the optical material density is proportional to the intensity envelope of the driving field~\cite{sommer2016attosecond,sommer2016ultrafast}. The increase in phase shift can be described as a linear rise with intensity $\Delta \phi_{NL}^0 = m\times I_0$, with $m = 1.06\times 10^{-13}$~cm$^2$/W. From the relation $m = kln_2$, where $k = 2\pi/\lambda$ and $l = 3.79 $ {\AA}, the propagation length, the nonlinear refractive index  $n_2 = 3.54 \times 10^{-11}$ cm$^2$/W can be extracted for 800 nm, 30 fs pulses. 

\subsection{Onset of dielectric breakdown}
\label{sec:optical_damage}

\begin{figure*}
	\centering{}
	\includegraphics[width=\textwidth, keepaspectratio]{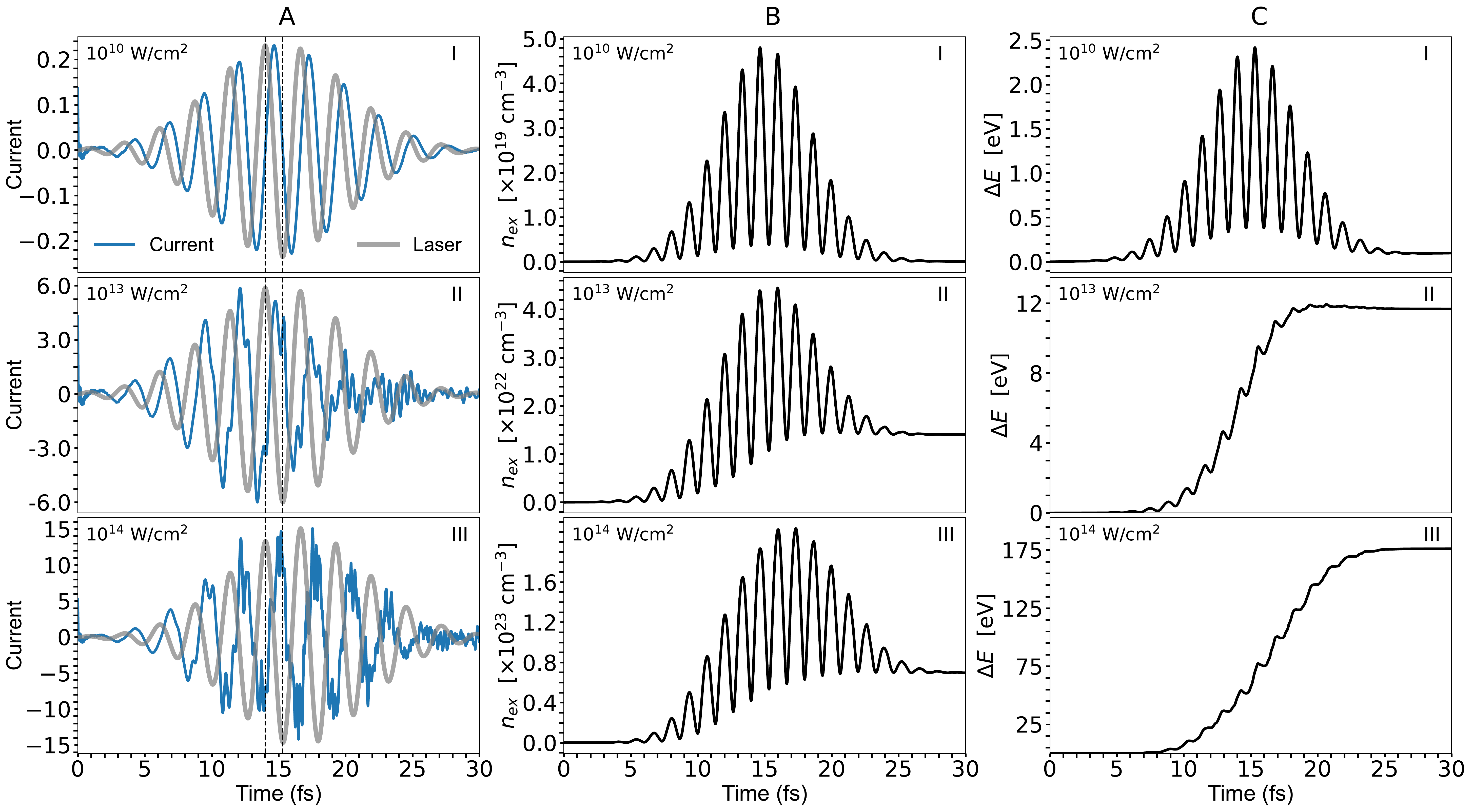}
    \caption{\label{fig:breakdown}Different regimes of the interaction of TiO$_2$ with 30 fs laser pulses at 800 nm with peak intensities $10^{10}$ W/cm$^2$ (top), $10^{13}$ W/cm$^2$ (middle) and $10^{14}$ W/cm$^2$ (bottom). Fig. A(I-III) displays the induced current density and electric field (scaled with respect to the current amplitude to show phase relations). Fig. B(I-III) shows the number density of excited electrons per cubic centimeter and Fig. C(I-III) represents the excitation energy. Dashed vertical lines in A(I-III) are given as a guide to the eye to show the phase variations.}
\end{figure*}

In Fig.~\ref{fig:breakdown}, we present the time evolution of current, energy, and excited electron density for three different peak intensities, $I_0 = 10^{10}, \ 10^{13}$ and $10^{14}$ W/cm$^2$. The laser frequency is $\omega = 1.55$ eV (800 nm), and the pulse duration is $T = 30$ fs. The time profiles of the electric field and the induced current are depicted in Fig.~\ref{fig:breakdown} A(I-III). The electric field profile is normalized with respect to the peak of the induced current at a given peak intensity to enable a comparison of the relative phase. Fig.~\ref{fig:breakdown} B(I-III) present the number of excited electrons calculated using Eq.~(\ref{eq:nex}), while Fig.~\ref{fig:breakdown} C(I-III) depicts the excitation energy defined in Eq.~(\ref{eq:energy_transfer}) as a function of time. 

The induced current at intensities $1\times 10^{10} $ W/cm$^2$ (Fig.~\ref{fig:breakdown} A(I)) follows the pulse's electric field with a phase shift of $\pi/2$, indicating a linear dielectric response. The excited electron density (Fig.~\ref{fig:breakdown} B(I)) and excitation energy (Fig.~\ref{fig:breakdown} C(I)) at this intensity oscillate synchronously with the electric field and the ground state conditions are restored after the interaction. The situation changes significantly at intensities of $10^{13}$ W/cm$^2$  and $10^{14}$ W/cm$^2$. The induced current during the interaction is distorted (Fig.~\ref{fig:breakdown} A(II) and A(III)), and the phase difference between the applied electric field and the induced current deviates from $\pi/2$. For $I = 1\times 10^{14} $ W/cm$^2$, the current and the electric field become nearly out-of-phase, indicating a strongly nonlinear response of electrons to the incident field~\cite{yamada2019energy}. Starting from about 10 fs, the number of excited electrons and the excitation energy increase rapidly at $10^{13}$ W/cm$^2$ (Fig.~\ref{fig:breakdown} B(II) and C(II)) and $10^{14}$ W/cm$^2$ (Fig.~\ref{fig:breakdown} B(III) and C(III)). By 20 fs, these quantities reach saturation values. Even after the laser pulse ends, the oscillation of the induced current persists, which is a clear indication of the onset of optical breakdown~\cite{otobe2008first}. This behavior is consistent with the abrupt increase in energy discussed in Sec.~\ref{sec:energy_transfer} due to resonant energy transfer at the breakdown. However, such oscillations will eventually decay due to dissipative processes such as electron-phonon coupling, impurity, and disorder scattering on longer time scales ($\gtrsim 100 $ fs)~\cite{wachter2014ab}.

Electrons excited into the conduction band exhibit a metallic response, resulting in collective plasmon mode. The plasma frequency corresponding to an electron density $n_{e}$ can be estimated by 
\begin{equation}
	\omega_p = \left(\frac{n_{e}e^2}{m\epsilon}\right)^{1/2} ,
\end{equation}
where $\epsilon$ is the dielectric constant of anatase TiO$_2$ ($\epsilon$ = 5.82)~\cite{gonzalez1997infrared}, $m$ and $e$ are the mass and charge of the electron respectively. 

At an intensity $1\times 10^{13}$ W/cm$^2$, the final number of excited electrons (Fig.~\ref{fig:breakdown} B(II)) is $1.4\times 10^{22}$ cm$^{-3}$. This corresponds to a plasma frequency of $\omega_p = 1.82 $ eV, slightly higher than the frequency of the applied laser pulse ($\omega_l = 1.55$ eV). As the intensity of the applied field increases, the density of electrons excited via the two-photon and tunneling mechanisms in the conduction band also gradually increases. When the electron density reaches a threshold where the plasma and laser frequencies are in resonance, a significant energy transfer occurs from the laser to the electrons. The low-amplitude coherent oscillations of the induced current observed on the trailing edge of the laser pulse (Fig.~\ref{fig:breakdown} A(II) and A(III)) results from the partial coherence between the involved non-stationary states left by the laser field. It is characteristic of plasmonic metal systems~\cite{zhang2020controlling}. This ultrafast and dissipative strong-field transition to plasmonic metal-like behavior is known as dynamic metallization~\cite{durach2011predicted,stockman2002coherent,stockman2007absolute}. Based on the dynamics presented in Fig.~\ref{fig:residual_energy} and in Fig.~\ref{fig:breakdown}, $I_0 = 1\times 10^{13}$ W/cm$^2$ can be identified as the intensity at which the laser-induced damage starts. For 30 fs pulses (11.7 fs FWHM), this intensity corresponds to a damage threshold of 0.1 J/cm$^2$.

\begin{figure*}
	\centering{}
	\includegraphics[width=0.8\textwidth, keepaspectratio]{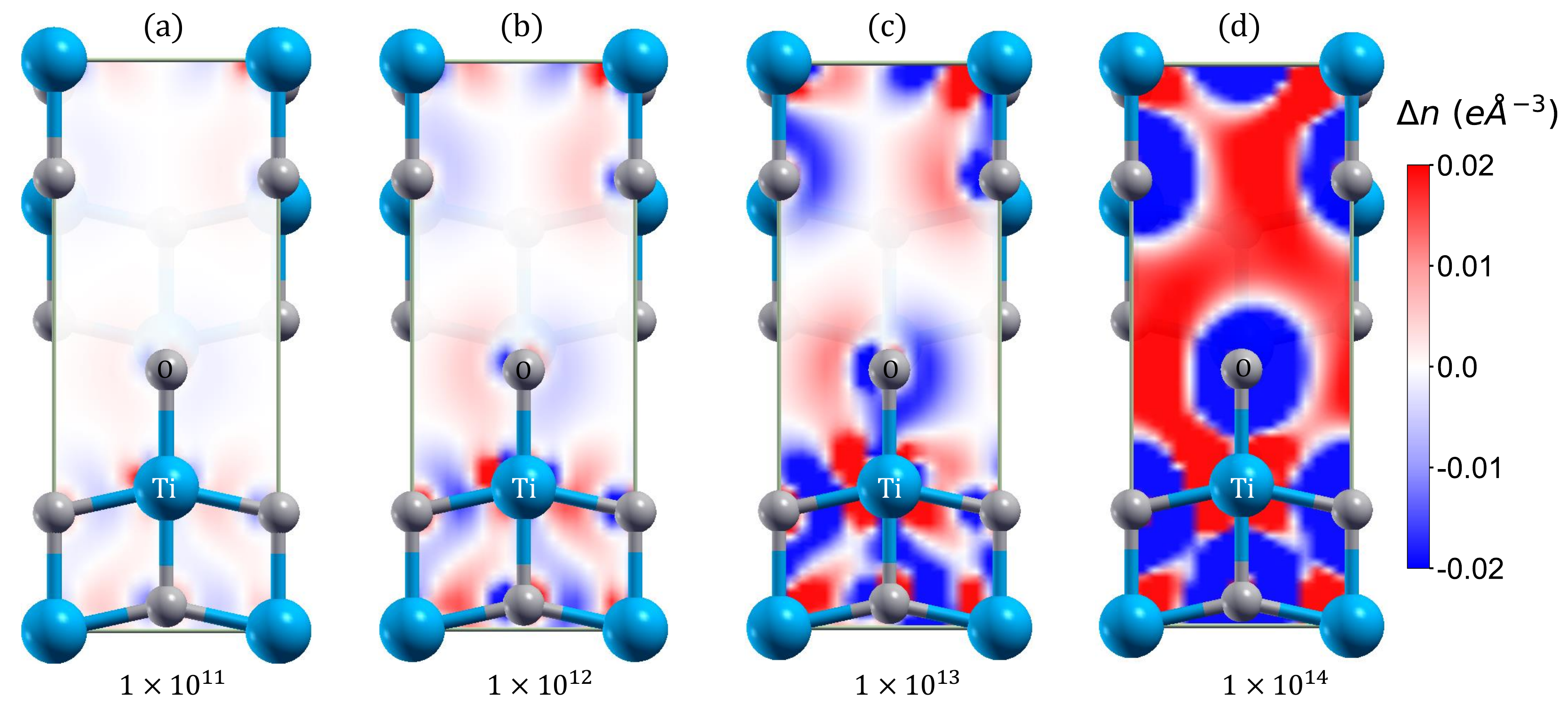}
    \caption{\label{fig:chg_difference}Snapshots of electron density difference with respect to the unperturbed state evaluated for different intensities. The snapshots displayed here for various peak intensities are taken at the same instant of time when the electric field of the pulse reaches zero right after its peak value. The red and blue colors indicate the gain and the loss of the density, respectively, with respect to the ground state.}
\end{figure*}

The dynamics outlined in the preceding section are represented by the change in electron density induced by the laser pulse  in Fig.~\ref{fig:chg_difference}. The snapshots displayed here for various peak intensities indicate the difference in electron density between the perturbed and unperturbed systems at the instant when the electric field of the pulse reaches zero right after its peak value. The positive (increase from the ground state) and the negative (reduction from the ground state) variations in the density are denoted in Figs.~\ref{fig:chg_difference} (c) and (d) by red and blue, respectively. When the laser is weak [Figs.~\ref{fig:chg_difference} (a) and (b)], the variation of electron density around ionic cores is uniform corresponding to a linear and adiabatic response. At higher laser intensity [Figs.~\ref{fig:chg_difference} (c) and (d)], the charge distribution extends into the interstitial region, indicating the laser-induced population of delocalized conduction band levels~\cite{zhang2017manipulation}.

\subsection{Comparison with experiments}
  \begin{table}[ht] 
	\caption{Summary of available experimental data for the nonlinear refractive index ($n_2$) of TiO$_2$ measured using ns and fs laser pulses at different wavelengths. The $n_2$ calculated in this work from TDDFT simulation is also given in the table for comparison.   }
\begin{ruledtabular}
	\begin{tabular}{*4c}
    $n_2$ (cm$^2$/W) & $\lambda$ (nm) & Pulse width & Ref.  \\
    \hline
      $\sim 10^{-14}$        & 532, 780 & 35 fs  & \cite{guo2017dependence}\\ 
      $6.32 \times 10^{-13}$ & 800      & 50 fs  & \cite{long2009third} \\ 
      $2.0\times 10^{-14}$   & 800      & 50 fs  & \cite{castillo2010ultrafast}\\ 
      $1.0\times 10^{-15}$   & 800      & 60 fs  & \cite{portuondo2008ultrafast}\\ 
      $2.5\times 10^{-11}$   & 800      & 250 fs & \cite{elim2003ultrafast} \\ 
      $6.2\times 10^{-11}$   & 800      & 250 fs & \cite{yuwono2004controlling}\\ 
      $1.2\times 10^{-13}$   & 532      & 5 ns   & \cite{watanabe1995measurement}\\ 
      $1.5\times 10^{-13}$   & 532      & 7 ns   & \cite{irimpan2008luminescence}\\ 
    \hline
      $3.54\times 10^{-11}$  & 800      & 30 fs  & This Work\\ 
		\end{tabular}
\end{ruledtabular}
\label{table:n2_comparison}
\end{table}
  \begin{table}[ht] 
	\caption{The measured value of the laser-induced damage threshold (LIDT) of TiO$_2$ available in the literature. The table also lists the LIDT calculated in the present work using TDDFT simulations.}
\begin{ruledtabular}
	\begin{tabular}{*4c}
    LIDT (J/cm$^2$) & $\lambda $(nm) & Pulse width & Ref.  \\
    \hline
      0.5   & 800      & 50 fs      & \cite{yao2008investigation} \\ 
      0.6   & 800      & 220 fs     & \cite{yao2008investigation}\\ 
      1.43  & 532      & 10 ns & \cite{kumar2020laser}\\ 
      2.09  & 1064     & 10 ns & \cite{kumar2020laser} \\ 
    \hline
      0.1   & 800      & 30 fs      & This Work\\ 
		\end{tabular}
\end{ruledtabular}
\label{table:lidt_comparison}
\end{table}

Now let's compare the figures estimated in the current work for TiO$_2$'s nonlinear refractive index ($n_2$) and laser-induced damage threshold (LIDT) with those found in the literature. The measured value of $n_2$ of TiO$_2$ reported in the literature is summarised in Table~\ref{table:n2_comparison}. The calculated value of $n_2$ for 30 fs pulses at 800 nm in the current work is about three orders of magnitude greater than that measured using identical wavelength and pulse widths~\cite{guo2017dependence}. The variability of experimental data of $n_2$ presented in Table~\ref{table:n2_comparison} shows that the duration and frequency of the laser pulse have a significant impact on the observed value of $n_2$. Moreover, the measured value of $n_2$ vary due to a variety of factors, including nonlinear refraction dispersion, different sizes and volume fractions of synthesized materials, the effect of structure confinement in the case of nanostructured compounds, etc.  The simulations described here are for the bulk phase of TiO$_2$, whereas the majority of the reported $n_2$ was for thin films of TiO$_2$. Additionally, the collisional relaxation not taken into account in the current work, becomes significant for laser pulses longer than $\approx 100$ fs. 

Table~\ref{table:lidt_comparison} presents the experimental literature for the laser-induced damage threshold (LIDT) of TiO$_2$. We calculated the damage threshold using the critical density criterion, similar to that measured in experiments with comparable parameters. The damage threshold depends on the frequency and duration of the laser pulse and dynamics toward thermal distribution. The thermal effects probably can be neglected in our case because an ultra-short pulsed laser ($<50$ fs) was used. The bandgap of TiO$_2$ in the present study is underestimated due to the GGA functionals~\cite{sruthil2022modified}. Using more accurate functionals leads to a larger bandgap. This would lead to a higher damage threshold in agreement with the trend of the experimental data~\cite{lee2014firstprinciples}.

\section{Summary}
\label{sec:conclusion}
We presented a systematic investigation of perturbative and non-perturbative electron dynamics of TiO$_2$ to 30 fs laser pulses at 400 nm and 800 nm using ab initio time-dependent density functional theory. The mechanism of nonlinear optical interaction of TiO$_2$ at different intensities is discussed. We can see the onset of laser-induced material damage and the accompanying plasmon dynamics from first-principles. The trends of the value of the nonlinear refractive index ($n_2$) and laser-induced damage threshold obtained from the simulations are consistent with the experimental data in the literature. Non-resonant, perturbative interactions at 800 nm and the accompanying nonlinear phase shift observed in TiO$_2$ well below the damage threshold hold promises incorporating TiO$_2$ in optical switches. The present study could guide the further exploration of laser parameters and structural and defect engineering of TiO$_2$ with tailored properties for specific applications, potentially leading to improved performance in nonlinear photonics devices. By pursuing these directions, researchers can advance the understanding and utilization of TiO$_2$ and similar materials for nonlinear photonics applications.

\section{Acknowledgments}
This work was in part supported by the Center for Advanced Systems Understanding (CASUS) which is financed by Germany's Federal Ministry of Education and Research (BMBF) and by the Saxon state government out of the State budget approved by the Saxon State Parliament.

\bibliography{references.bib}

\clearpage

\section*{Supplemental Material}

\begin{figure}[ht]
  \centering{}
  \includegraphics[width=0.8\textwidth, keepaspectratio]{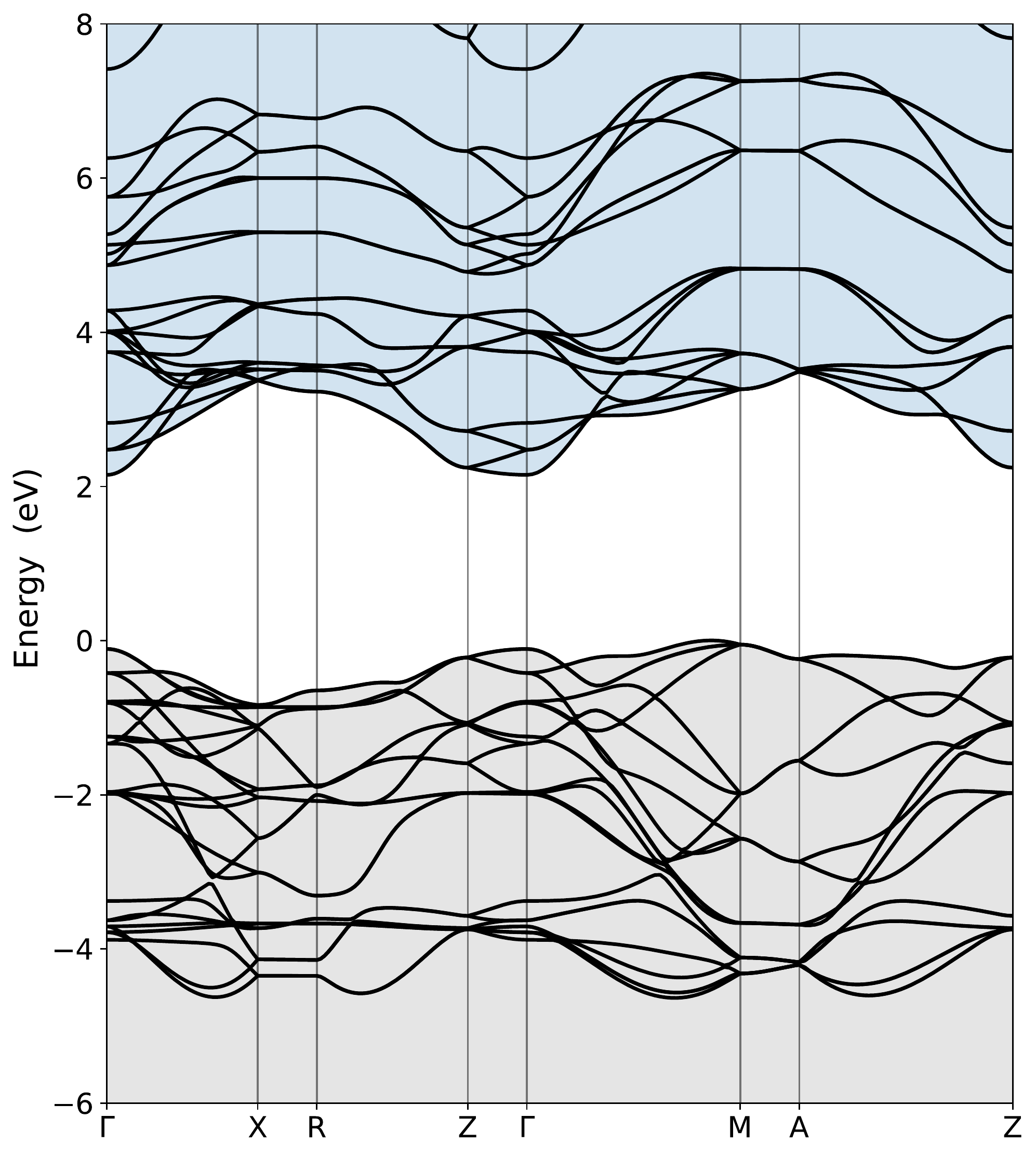}
  \caption{The GGA (PBE) electronic bandstructure of anatase TiO$_2$. We obtain an indirect band gap of 2.15 eV and a direct band gap of 2.25 eV. }
  \label{fig:pbe_bands_tio2}
\end{figure}

\begin{figure}[ht]
  \centering{}
  \includegraphics[width=0.8\textwidth, keepaspectratio]{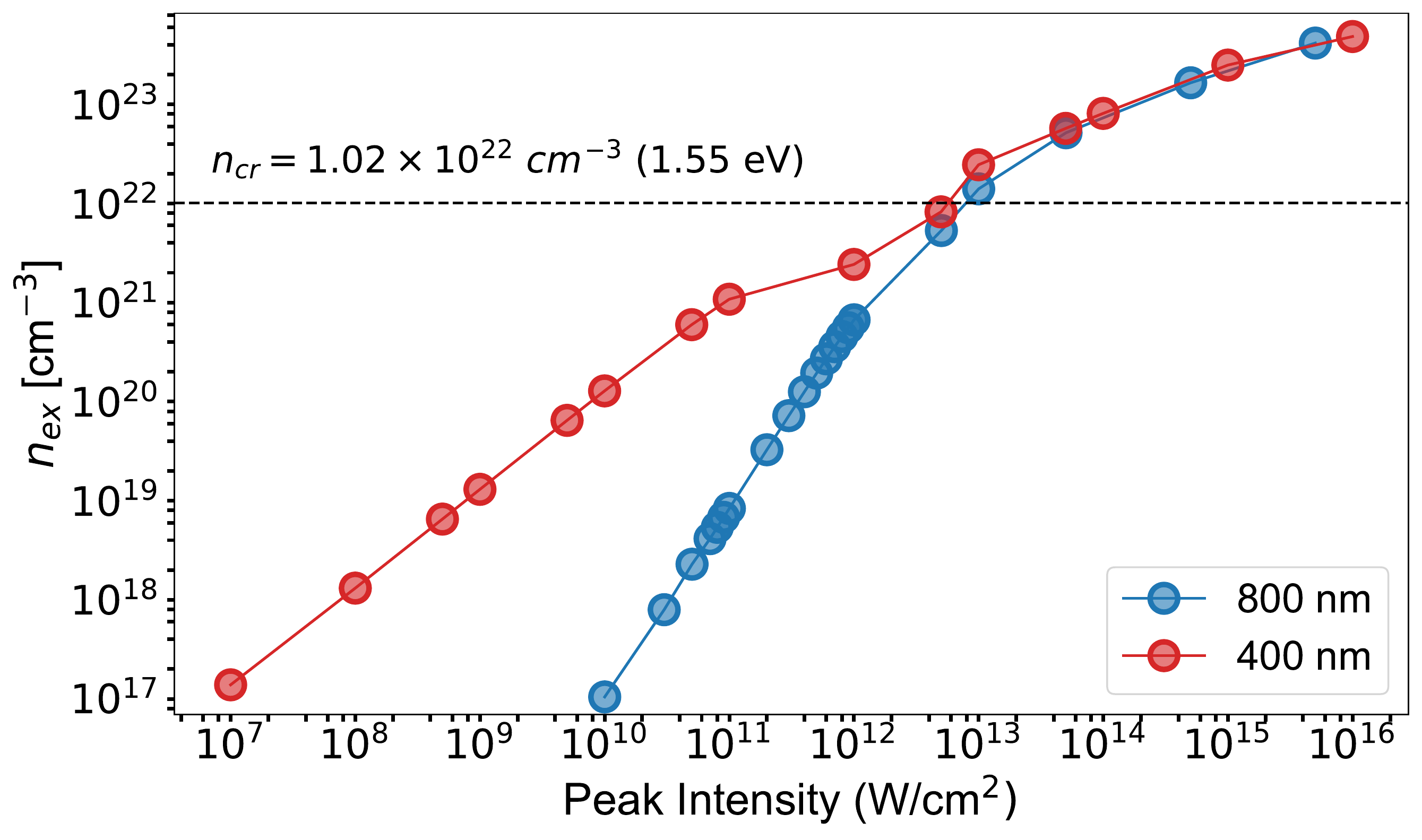}
  \caption{The number of excited electrons in TiO$_2$ at their saturation values created when interacting with 800 nm and 400 nm laser pulses at different peak intensities. The critical density corresponding to 1.55 eV (for 800 nm pulses) is also marked in the figure with a horizontal dashed line.}
  \label{fig:n_exc_800_400}
\end{figure}

\begin{figure}[ht]
  \centering{}
  \includegraphics[width=\textwidth, keepaspectratio]{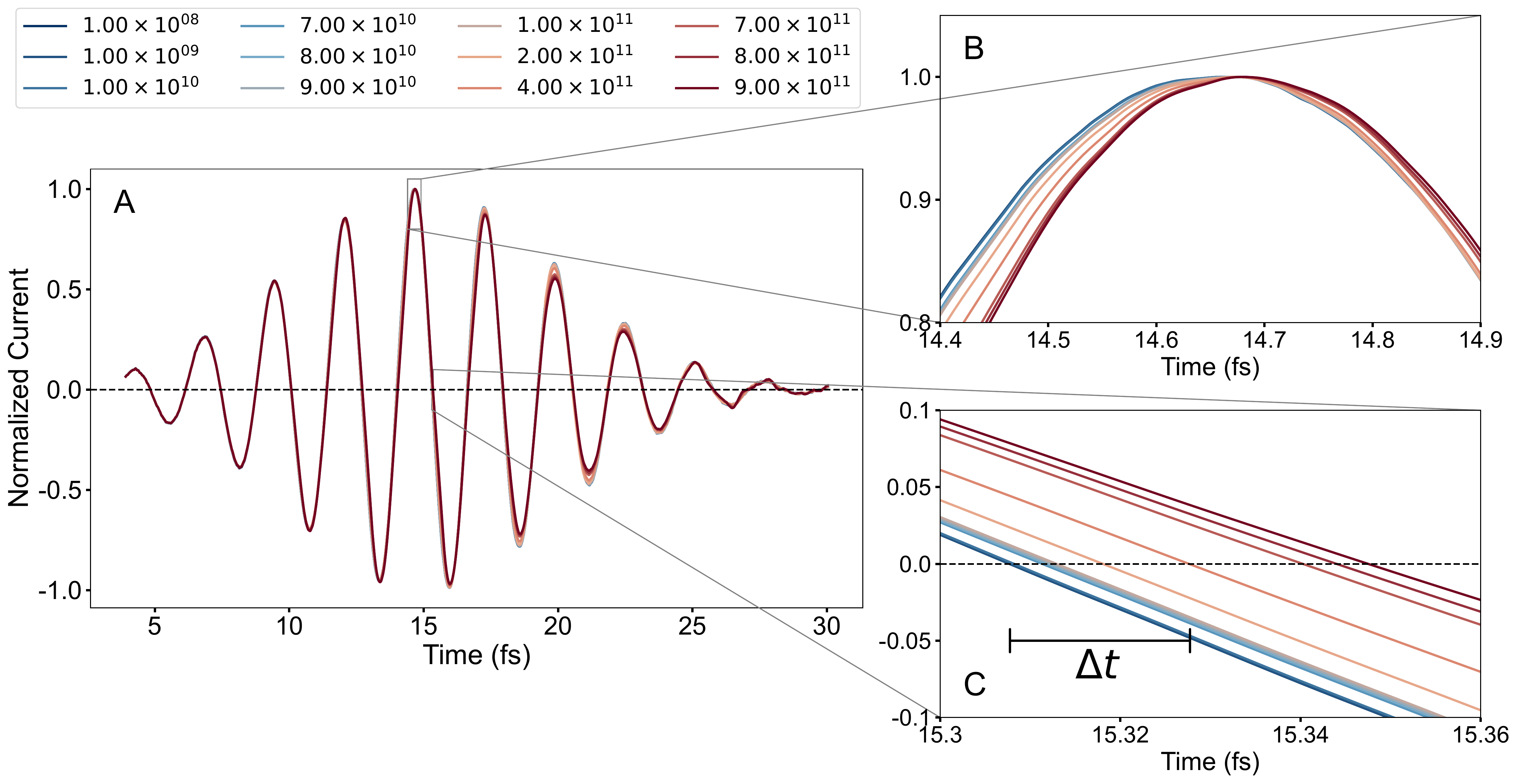}
  \caption{The phase shift of photoinduced current accumulated during the interaction of TiO$_2$ with 800 nm pulses of different peak intensities is shown. Panel A shows the photoinduced current profiles for different laser peak intensities. The inset panel B shows the delay of the peaks of the photoinduced current at different intensities with respect to that at $1\times 10^8$ W/cm$^2$. The inset panel C shows the time delay of the photoinduced current when it crosses zero immediately following the peak value. The time delay at these zero-crossing locations is used to estimate the phase shift and corresponding change in the nonlinear refractive index ($n_2$) of TiO$_2$}
  \label{fig:nl_phase_shift}
\end{figure}

\end{document}